# Finite Future Cosmological Singularity Times and Maximum Predictability Times in a Nonlinear FRW-KG Scalar Cosmology

John Max Wilson
Keith Andrew

Department of Physics and Astronomy
Western Kentucky University
Bowling Green, KY 42101

Abstract

We investigate the relative time scales associated with finite future cosmological singularities, especially those classified as Big Rip cosmologies, and the maximum predictability time of a coupled FRW-KG scalar cosmology with chaotic regimes. Our approach is to show that by starting with a FRW-KG scalar cosmology with a potential that admits an analytical solution resulting in a finite time future singularity there exists a Lyapunov time scale that is earlier than the formation of the singularity. For this singularity both the cosmological scale parameter a(t) and the Hubble parameter H(t) become infinite at a finite future time, the Big Rip time. We compare this time scale to the predictability time scale for a chaotic FRW-KG scalar cosmology. We find that there are cases where the chaotic time scale is earlier than the Big Rip singularity calling for special care in interpreting and predicting the formation of the future cosmological singularity.

## I. Introduction

The field of chaotic dynamical systems has grown to include a wide class of known physical phenomena in both the fundamental theoretical domain and within the context of analyzing empirical data. These studies have given rise to a large number of general methods and specific tools for analyzing nonlinear dynamics[1]. Many of these methods have been used extensively to investigate a number of interesting nonlinear cosmological models. In particular the Friedman-Robertson-Walker (FRW) universe coupled to a massive scalar Klein-Gordon (KG) field is known to exhibit chaotic behavior for certain values of the fields and their potentials [2-5]. In these models the chaos can manifest itself as an infinite series of bounces for the scale factor of the FRW universe, a(t). In the chaotic regime the scale factor trajectory is a fractal and exhibits all the hallmarks of a chaotic system and may be characterized as such by its topological entropy [6,7]. Early work often focused on avoiding a future inevitable cosmological collapse singularity[8,9], a big crunch, by replacing the smooth evolution of the scale factor with this underlying fractal pattern by a judicious choice of parameters and/or initial conditions[10]. One of the first papers to examine this type of cosmology is due to Page [11] where he used a quadratic scalar field potential in the KG equation. This work indicated that the dynamical behavior of this nonintegrable model is similar to the chaos appearing in ergodic systems. Since then, the problem of chaos in FRW scalar field cosmologies has been investigated thoroughly in several papers. The dynamical properties of such cosmological models are often studied in the context of dynamical systems exhibiting chaotic scattering. These authors used an index of chaoticity which is capable of detecting stochastic regions in the scattering version of this problem. In this way, one can attempt to compare theoretical predictions with observational data by utilizing the measured evolution of the Hubble parameter H [12-16].

As the class of acceptable potential functions for the fields has enlarged a more robust set of final states has been discovered [17-20], and as Toporinski has pointed out, even the Damour Mukhanov potential[15] with the scalar field power q as low as 1.2 has been proven to be strongly chaotic. Applications of such studies have been extended by investigating the properties of multiple scalar or supersymmetric fields and in searching for low energy limits to string and brane theory solutions and early inflation[17-19]. In particular using potentials that decay faster than the scalar field quadratic potential, i.e. an exponentially decaying potential function for the scalar field potential, gave rise to the discovery that there are regions where the chaos will vanish and the solutions are perfectly regular and remain deterministic[14, 15]. These potentials exhibit a transition region for certain parameters where the system can leave the chaotic regime and never return. Even with this larger class of potentials many models still give rise to a distant future singularity. However for systems that remain in the chaotic regime the notion of predictability is severely limited, one can not calculate measurable parameters to arbitrarily distant times with certainty. For all chaotic systems however, there is a maximum predictability time beyond which the inherent nonlinear chaos has erased any chance of accurate prediction. An estimate of this time can be found by using the inverse of the positive Lyapunov exponents, known as the Lyapunov timescale[1,5]. This concept can be generalized to a more accurate and general measure of the maximum predictability time for a chaotic system that can be represented by a set of nonlinear differential equations while being careful of inherent instabilities. Such a treatment was developed by Liao who has considered the time scale associated with a loss of information due to a sensitivity to initial conditions and generalized the definition of maximum predictability time to include the growing loss of predictability of a future prediction caused by the inaccuracy of the initial conditions, repeated rounding errors or fundamental constants in the theory[20,21].

Considerable progress on the FRW-KG cosmology has also been made using a geometric description of chaos in Hamiltonian systems of cosmological origin using the tools of pseudo-Riemannian geometry [13, 22]. The approach using these methods is to analyze geodesic flows in the appropriate metric- in this case the Jacobi metric. Chaos can manifest itself by curvature invariants that become singular indicating a sensitive dependence on initial conditions and by spacelike geodesics that change to timelike geodesics countably many times or even infinitely many times. For the case of conformally coupled scalar fields the nonlinear dynamics classification has been carried out and the Lyapunov exponents calculated, a powerful general theorem may soon be forthcoming in this area while the special tools currently available are widely used[22].

As these scalar-FRW models were widened to include possible new effects due to the recently discovered cosmic dark energy a more general equation of state has been invoked[23,24]. In first order models this is characterized by having a density and a pressure that are proportional to each other: $\rho = w\, p$ where w is the equation of state parameter measured to be near[25] -1. This allows for an interpretation of the coincidence problem[3, 26] while the negative dark energy pressure is responsible for the accelerating expansion of the scale factor and often leads to a future singularity. A classification of FRW singularities was carried out by Fernandez Jambrina[27] and Cattoen and Visser[28] . These authors considered a general class of singular states that are kinematical and dynamic and find the necessary and sufficient conditions for both spacelike and timelike singularities considered as big bangs, bounces, crunches, rips, sudden singularities and other extremality events that can lead to undefined values of curvature invariants or nonextensibility of geodesic curves. The literature on finite future big rip singularities is extensive and includes avoidance of the big rip[29], traveling around a big rip with a wormhole, referred to as a big trip[30,31], extended gravity singularities in terms of Gauss Bonnet[32], higher derivative[33] , Casimir energy[34], f(R) extensions[35,36], effects of quantization[37], impact of renormalization[38] and a class of pseudo rip models[39,40]. Here we will refer to a subset of such singular solutions using the future singularities recognized by Frampton and Scherer[39] who identified three essential future singularities based upon the behavior of the Hubble parameter H: 1. Big Rip: a finite future time singularity where H→ ∞ and $t_{rip} < \infty$, 2. Little Rip: H →∞ and t→∞, 3. Pseudo Rip: H→ constant, and t→∞. A small set of these models are analytically solvable and future singularity times can be estimated given the current values of the cosmological parameters. For reasonable values of the cosmological parameters the class of finite future cosmological singularities that are stable[41] and yield singularity times are generally found to be quite distant compared to the current age of the universe.

In this paper we will give an example showing that for a general phantom energy FRW-KG scalar cosmological model which is chaotic and exhibits a finite future singularity the Lyapunov predictability time can proceed the finite future singularity formation time and thereby bring into question whether or not the predicted singularity is physically meaningful. The structure of the paper is as follows: we first present the FRW-scalar field equations and a typical solution structure for a dark energy scenario, in the next section we introduce a conventional predictability time measure, this is followed by an explicit example and we present our conclusions in the last section.

**II. FRW-Scalar Cosmology with a Big Rip**

In geometric units we begin with the action

$$S = \int d^4x \sqrt{-g}\left\{\frac{m_P^2}{16\pi}R + \frac{1}{2}g^{\mu\nu}\partial_\mu\varphi\,\partial_\nu\varphi - V(\varphi)\right\} \quad (1)$$

for scalar curvature R, scalar field φ, and potential function V(φ). Applying the FRW metric ansatz and definition for the Hubble constant H,

$$ds^2 = dt^2 - a^2(t)d^2\Omega^{(3)}$$
$$H(t) = \frac{\dot{a}}{a} \quad (2)$$

to a variation of the action gives rise to two second order equations

$$\frac{m_P^2}{16\pi}\left(\ddot{a}+\frac{\dot{a}^2}{2a}+\frac{1}{2a}\right)+\frac{a\dot{\varphi}^2}{8}-\frac{aV(\varphi)}{4}=0$$
$$\ddot{\varphi}+\frac{3\dot{\varphi}\dot{a}}{a}+\frac{\partial V}{\partial \varphi}=\ddot{\varphi}+3H\dot{\varphi}+\frac{\partial V}{\partial \varphi}=0$$ (3)

where the scalar field dynamical equation behaves like an oscillator with a time dependent friction given by 3H for V(φ) a smooth nonnegative function. These equations represent a type of system of nonlinear coupled oscillators. One of the most important oscillator applications in modern cosmology corresponds to the slow-roll inflation regime which can occur when the "friction" term is much larger than the frequency of the oscillator. More precisely, the slow-roll approximation is characterized by the system

$$H=\sqrt{\frac{8\pi}{3m_P^2}V(\varphi)}\quad \dot{\varphi}=\frac{V'(\varphi)}{3H}$$ (4)

which is arrived at by neglecting the second derivative terms, the kinetic energy of the scalar field and the spatial curvature. This regime is rather natural for physically admissible initial conditions [5, 14, 15] and leads to fast growth of the scale factor a(t) while the scalar field φ slow rolls toward zero. When the scalar field φ falls below some value φ ~ $m_p$ this regime disappears[16].

In the opposite case, when the "friction" is small, the dynamics of φ is characterized by damped oscillations, this regime is typical for late time evolution of the Universe. However, in contrast to zero- or negative spatial curvature cases, for a Universe with a positive spatial curvature this regime does not characterize the final sate of the system which is ultimately followed by a recollapse of the Universe. Unlike a recollapse, a transition from contraction to expansion (often called a "bounce") is also possible, which requires specially imposed initial conditions[5,6,12,14]. These two characteristic features of a positive spatial curvature case – ultimate recollapse and the existence of initial conditions, leading to a bounce – result in a complicated dynamics which in some situations may be chaotic. A third case, and the one of interest here, is also possible, a Big Rip scenario characterized by a finite future singularity with no intermediate bounce. [42-48] These equations also yield a first integral of the motion

$$-\frac{3}{8\pi}m_P^2(\dot{a}^2+1)+\frac{a^2}{2}(\dot{\varphi}^2+2V(\varphi))=0$$ (5)

corresponding to the identically zero Hamiltonian. From Eq.(5) the region where $\dot{a}=0$, corresponding to extrema in contraction or expansion, is given by

$$a^2 \leq \frac{3}{8\pi}\frac{m_P^2}{V(\phi)}$$ (6)

where the equality defines the "Euclidean" boundary. Page, in 1983[11], found resonant chaotic behavior in a cosmological model describing the evolution of a closed Friedman–Robertson–Walker (FRW) Universe filled with a massive scalar field φ(t) as described in Eqs.(3). The analog of scattering on a disc in his model is a "bounce" – a transition from a cosmological collapse to a cosmological expansion of the Universe. The final regime of the dynamics for almost all initial conditions is falling into a cosmological singularity. The equivalent description of the set of periodic trajectories in this model in the language of symbolic dynamics and in the calculation of topological entropy which has been carried out[10,13]. Later, such an analysis was done for cosmological models with other types of a scalar field[5, 14, 16]. It appears that, depending on the particular form of the scalar field potential, the dynamics may be either chaotic or regular. Several types of transitions from chaos to a non-chaotic dynamics for particular one-parameter families of potential were described in detail by Toporensky[5].

Phantom energy allows for the possibility to describe primordial inflation[47]-it is known to possess the following unusual characteristics. If dark energy is described by a scalar field φ with the FRW customary definitions, $\rho = -\dot{\phi}^2/2 + V(\phi)$  $p = -\dot{\phi}^2/2 - V(\phi)$, with ρ and p the energy density and pressure, respectively, and V (φ) the field potential, then (i) the kinetic term is less than zero: $\dot{\phi}^2/2 < 0$ and therefore phantom cosmologies suffer from violent instabilities and classical inconsistencies, (ii) the energy density is an increasing function of time which would make the quantum-gravity regime critical to understanding later times, (iii) the dominant energy condition is violated so that ρ+p < 0, (iv) there will be a singularity in the finite future often known as the Big Rip at which the universe ceases to exist and (v) near the Big Rip singularity there may appear cosmic violations of causality. These properties correspond to the definition of phantom energy in the quintessence scenario.

A Big Rip scenario for a minimally coupled scalar field for a phantom energy EoS where the pressure and Hubble relations are given by

$$w = \frac{p}{\rho} = \frac{-\dot{\phi}^2 - 2V(\phi)}{-\dot{\phi}^2 + 2V(\phi)} \quad H = \frac{\dot{a}}{a} \quad w = -1 - \xi \quad \xi > 0$$
$$\frac{\dot{\rho}}{\rho} = -3H(1+w) = \frac{2\dot{H}}{H} \quad \frac{a}{\rho}\frac{\partial \rho}{\partial a} = -3(1+w)$$

(10)

that corresponds to the solution of Eqs.(3) has been found by Yurov[29] where the scale factor a(t) and the density are given by

$$a(t) = \left[a_o^{3(1+w)/2} + \tfrac{3}{2}C(1+w)(t-t_o)\right]^{\frac{2}{3(1+w)}}$$

$$\rho(t) = H^2 = \frac{C^2}{\left[a_o^{\frac{3(1+w)}{2}} + \tfrac{3}{2}C(1+w)(t-t_o)\right]^2} \tag{11}$$

$$C = \frac{6a_o^{-\frac{3}{2}(w-1)}\left(1 + a_o^{3w}H\xi + w\right)}{w-1}$$

and the scalar potential and scalar field solution is

$$V(\phi) = \tfrac{1}{2}(|w|+1)C^2 e^{-3i(\phi-\phi_o)\sqrt{|w|-1}}$$

$$\phi(t) = \phi_o - \frac{2i}{3\sqrt{|w|-1}}\log\left[\left(\frac{a_o}{a(t)}\right)^{3(|w|-1)/2}\right] \tag{12}$$

Leading to a finite future singularity time for the Big Rip at

$$t_{rip} = \frac{2a_o^{\frac{-3(|w|-1)}{2}}}{36a_o^{-\frac{3}{2}(w-1)}\left(1+a_o^{3w}H\xi+w\right)} + t_o. \tag{13}$$

This model clearly possesses a finite future singularity which can be arbitrarily close to $t_0$ depending upon the value of w. This singularity is displaced but not removed if we expand the potential in a Taylor series keeping the self interacting terms. For the quadratic and cubic terms the singularity formation time remains analytically solvable. This allows for a comparison between the Big Rip formation time to the predictability time in a chaotic region. To examine a Big Rip that occurs near a chaotic regime we use the results of Toporensky[5,14,15] where the analytical criterion on the steepness of the potential is used to determine if there is chaos, this occurs whenever the scale factor, a, and field φ satisfies the inequality

$$\frac{\ddot{\varphi}}{\ddot{a}} < \frac{d\varphi}{da} \tag{14}$$

In terms of the potential this condition on the Euclidean boundary is

$$V(\varphi) > \sqrt{\frac{3m_P^2}{16\pi}\frac{dV}{d\varphi}} \tag{15}$$

which is a restriction on the steepness of the potential. Any cosmology that satisfies this criterion at the Euclidean boundary can exhibit chaos, if they are steeper than this none of the trajectories will exhibit chaotic behavior.

## III. Critical Predictability Time $T_c$

Following Liao[20,21] we recognize that the critical predictability time $T_c$ has been defined in several different ways. A common numerical method presented in detail by Teixeira[42], is to take a numerical result given by the smallest time-step as *assumed* to be closest to the exact solution. Then Teixeira defined $T_c$ by means of a state vector on an $L^2$ normed error space to be between the result obtained by the smallest time-step and the result by some given larger one. This kind of definition includes the error induced by each time-step and thus is a global one for decoupling the trajectories. However, the decoupling of two curves is essentially a local occurrence. Thus, Liao gives a local definition of critical predictability time $T_c$, which is based on the geometrical characteristic of decoupling of two trajectories valid for any two nearby curves. Mathematically, let $a_1(t)$ and $a_2(t)$ denote two trajectories given by different nearby trajectories for a given dynamical system. Then the critical predictability time $T_c$ for two nearby cosmic scale factor trajectories $a_1(t)$ and $a_2(t)$, and two small constants $\varepsilon$ and $\delta$, can be determined by the following criteria:

$$a_1(t) = a(t_1), a_2(t) = a(t_2) = a(t_1 + \Delta t) \quad \dot{a} = \frac{da}{dt}$$

$$0 < t < T_C \quad \dot{a}_1 \dot{a}_2 < -\varepsilon, \quad \left|1 - \frac{a_1}{a_2}\right| > \delta, \quad t = T_C \qquad (16)$$

$$\varepsilon > 0, \quad \delta > 0$$

for a given $\varepsilon$, here selected to be ~0.5, the time t may be continued until $\delta$ is reached, here tested at 10%. The critical predictability time $T_c$ can then be interpreted as follows: the influence of truncation error, round-off error, nonlinear deviation and inaccuracy of initial conditions on numerical solutions is negligible in the interval $0 < t < T_c$, so that the computed result is predictable and thus can be regarded as a reliable solution in this interval. Eventually the ratio of the two trajectories will deviate significantly for a chaotic system forcing $\delta$ to grow uncontrollably for any $\varepsilon$. Using this concept of the critical predictability time $T_c$, the well known statement that "accurate long-term prediction of chaos is impossible" can be more precisely expressed as "accurate predictions of chaos beyond the critical predictability time $T_c$ is impossible." Here, $T_c$ is regarded as a critical point: computed results beyond the critical predictability time Tc are unreliable and should not be interpreted as providing physical insight about the dynamics of the system. Thus, the critical predictability time $T_c$ provides a strategy to detect the reliable part from a given solution. As pointed out by Lorenz[49], computational chaos (CC), computational uncertainty (CU) and computational periodicity (CP) are mainly based on the evaluation of Lyapunov exponents[50], which is a long-term property, any computed results for $t > T_c$ are unreliable, and thus have no direct physical meaning. Here we have an explicit form of a(t) and use this as a check of $T_c$ compared to the predictability time given by the Lyapunov exponent.

To calculate the Lyapunov exponent near an equilibrium point the system of equations is expanded about the equilibrium point and linearized:

$$\frac{d\vec{x}}{dt} = \vec{F}(\vec{x},t) \to \frac{d\vec{x}}{dt} \approx \hat{D}_J \vec{F}(\vec{x}_o,t) \cdot (\vec{x}-\vec{x}_o)$$

$$\hat{D}_J \vec{F} = \hat{J}_F = J\left(\frac{\partial F_i}{\partial x_j}\right) = \begin{pmatrix} \frac{\partial F_1}{\partial x_1} & \cdots & \cdots & \frac{\partial F_1}{\partial x_N} \\ \frac{\partial F_N}{\partial x_1} & \cdots & \cdots & \frac{\partial F_N}{\partial x_N} \end{pmatrix}\Bigg|_{\vec{x}=\vec{x}_o} \quad . \tag{17}$$

The FRW-KG massive scalar field system can be rewritten as a system of four first order nonlinear ODEs by using the first integral of the motion and defining a new field given as the time derivative of the field φ yielding four equations of the form:

$$\begin{aligned}
\dot{a} &= x \\
\dot{\phi} &= y \\
\dot{x} &= \frac{aV(\phi)}{4b} - \frac{x^2+1}{2a} - \frac{ay}{8b} \\
\dot{y} &= -\frac{3xy}{a} - V'(\phi) \\
\text{where} \quad b &= \frac{m_p^2}{16\pi} \quad \text{and} \quad V'(\phi) = \frac{\partial V}{\partial \phi} \quad .
\end{aligned} \tag{18}$$

Equilibrium values for maximum future expansion are given by

$$x = y = 0, \quad \phi = -\frac{i}{3\sqrt{\xi}}, \quad a = \pm\frac{\sqrt{8b\xi}\exp\left(-\frac{3}{2}i\phi_o\sqrt{\xi}\right)}{c\sqrt{|w|^2-1}}$$

$$\text{where} \quad c = \frac{6a_o^{-\frac{3}{2}(w-1)}\left(1+a_o^{3w}H\xi+w\right)}{w-1} \quad . \tag{19}$$

Using the expansion for the potential for the system of Eq.(12) and expanding around these equilibrium values we arrive at the largest Lyapunov exponents as

$$\lambda_{\pm} = \frac{\pm 36 a_o^{-\frac{3}{2}(w-1)}\left(1 + a_o^{3w} H\xi + w\right)\sqrt{e^{3i\phi_o\sqrt{|w|-1}}\left(|w|^2 - 1\right)}}{(w-1)\sqrt{2}} \quad .$$

(20)

For the case of real nondegenerate negative solutions we have an attractor, an unstable critical point if positive, and a saddle point for opposite signs. The Lyapunov exponents measure the rate of separation of trajectories in phase space. Only if trajectories separate exponentially fast do they have positive exponents. Systems with positive Lyapunov exponents are said to exhibit a sensitive dependence on initial conditions - one of the two ingredients of chaos (the other being the mixing and folding of trajectories). The inverse of the largest positive Lyapunov exponent is often referred to as the Lyapunov timescale. This timescale sets the dynamical timescale over which chaotic effects make themselves felt. In general relativity, Lyapunov exponents must be used with extreme care, if at all, as they are coordinate dependent. Indeed, a simple coordinate transformation can give a non-chaotic system positive exponents and a chaotic system vanishing exponents, as a caution we compare these to the critical predictably time of Eq. (16). From the largest positive eigenvalue we have the Lyaponuv predictability time of

$$t_{pred} = \frac{a_o^{\frac{3}{2}(w-1)}(w-1)}{9\sqrt{2}\left(1 + a_o^{3w} H(w-1) + w\right)\sqrt{e^{3i\phi_o\sqrt{|w|-1}}\left(|w|^2 - 1\right)}} - t_o$$

(21)

where we have used Eq.(18) to simplify the final expression. Comparison with Eq.(13) shows that for some field values the singularity formation time is later than the predictability time. This is the case where the formation of the singularity needs to be treated with caution.

**Conclusion**

We have seen that it is not difficult to construct a reasonable FRW-KG model that is both chaotic and exhibits a finite future singularity of the Big Rip type with a phantom energy equation of state that mimics dark energy. However when the time scale for chaos proceeds the formation of the future singularity the interpretation of the formation of the singularity and long term history of the universe must be done with great care. Both the critical predictability time and the Lyanpunov time depend upon the value of w, the more negative the value consistent with the phantom energy model the earlier the predictability time. Using the tools from geometric chaos theory we note that the curvature invariants are also singular and geodesic flows with timelike and spacelike properties oscillate back and forth from one into the other blurring the lightcone structure of the pseudo Riemannian spacetime manifold which provides the underlying causal structure. This means that the singularity time may be significantly displaced from the given theoretical value or that it may be avoided altogether. By the same token due to the sensitive dependence on initial conditions it may be that models that do not indicate the existence of any type of future singularity may indeed have the universe end in this fashion.